%% file: thickli_v4.tex
\documentclass[structabstract]{aa}  
\usepackage{natbib}
\usepackage{graphicx}
\usepackage{url}

\usepackage{rotating}

\usepackage{txfonts}
\usepackage[T1]{fontenc}
\usepackage[latin9]{inputenc}
\begin{document}

\def\teff{$T\rm_{eff }$}
\def\kms{$\mathrm {km~s}^{-1}$}
\def\gtsima
{\hbox{\raise0.5ex\hbox{$>\lower1.06ex\hbox{$\kern-1.07em{\sim}$}$}}}
\def\ltsima
{\hbox{\raise0.5ex\hbox{$<\lower1.06ex\hbox{$\kern-1.07em{\sim}$}$}}}

\title{Lithium-rich giants in the Galactic thick disk\thanks{Based on
observations taken at the Las Campanas and La Silla/Paranal observatory 
(ESO proposal ID: 077.B-0348)}}

   \author{L. \,Monaco\inst{1}, 
S. \,Villanova \inst{2},
C. \,Moni Bidin \inst{2},
G. \,Carraro\inst{1}, 
D. \,Geisler \inst{2},
P. \,Bonifacio\inst{3}, 
O.~A. \,Gonzalez\inst{4},
M. \,Zoccali\inst{5}
\and 
L. \,Jilkova\inst{1,6}
          }

\institute{
European Southern Observatory, Casilla 19001, Santiago, Chile
\and
Universidad de Concepci\'on,
Casilla 160-C, Concepci\'on, Chile
\and
GEPI, Observatoire de Paris, CNRS, Universit\'e Paris Diderot ; Place Jules
Janssen, 92190 Meudon, France
\and
ESO, Karl-Schwarzschild-Strasse 2, D-85748 Garching, Germany
\and
Departamento Astronom{\'\i}a y Astrof{\'\i}sica, Pontificia Universidad 
Cat\'olica de Chile, Av. Vicu\~na Mackenna 4860 Stgo., Chile 
\and
Department of Theoretical Physics and Astrophysics, Faculty of Science,
Masaryk University, (Kotl\'{a}\v{r}sk\'{a} 2, CZ-611 37) Brno, Czech
Republic
}

\authorrunning{Monaco et al.}
\mail{lmonaco@eso.org}

\titlerunning{Li-rich RGB stars in the Galactic thick disk}

\date{Received ...; Accepted...}

\abstract
{Lithium is a fragile element, which is easily destroyed in the stellar
interior. The existence of lithium-rich giants still represents a challenge for
stellar evolution models.}
{We have collected a large database of high-resolution stellar spectra of 824 
candidate thick-disk giants having 2\,MASS photometry and proper motions
measured by the Southern Proper-Motion Program (SPM). In order to investigate
the nature of Li-rich giants, we searched this database for giants presenting a
strong Li\,I resonance line.} 
{We performed a chemical abundance analysis on the selected stars with
the MOOG code along with proper ATLAS-9 model atmospheres. The iron content and
atmospheric parameters were fixed by using the equivalent width of a sample of
Fe lines. We also derive abundances for C, N, and O and measure or derive lower
limits on the  $^{12}$C/$^{13}$C isotopic ratios, which is a sensible diagnostic
of the stars evolutionary status.} 
{We detected five stars with a lithium abundance higher than 1.5, i.e.
Li-rich according to the current definition. One of them (SPM-313132) has
A(Li)$>$3.3 and, because of this, belongs to the group of the rare super Li-rich giants.
Its kinematics makes it a likely thin-disk member and its atmospheric parameters
are compatible with it being a 4\,M$_\odot$ star either on the red giant branch
(RGB) or the early asymptotic giant branch. This object is the first super 
Li-rich giant detected at this phase. The other four are likely low-mass thick-disk 
stars evolved past the RGB luminosity bump, as determined from their
metallicities and atmospheric parameters. The most evolved of them lies close to
the RGB-tip. It has A(Li)$>$2.7 and a low $^{12}$C/$^{13}$C isotopic ratio,
close to the cool bottom processing  predictions.}{}


\maketitle

\section{Introduction \label{sec:Introduction}}

Lithium is a fragile element, which is easily destroyed in the stellar interiors
at temperatures higher than $\sim$2.5$\times$10$^6$\,K. As soon as a star
evolves to the red giant phase, the convective envelope deepens and bring to the
stellar surface material that has been exposed to high temperatures in the
stellar interior, causing an overall dilution of Li in a giant's atmosphere. Red
giant stars are, therefore, expected to present a low amount of Li. Starting
from the meteoritic abundance of A(Li)\footnote{A(Li)=log
$\frac{n(Li)}{n(H)}$+12.}=3.3\,dex \citep[][]{grevesse98}, standard models
predict the dilution of Li in the giants' atmosphere down to a level of about
1.5\,dex, depending on the stellar mass and metallicity \citep[][]{ibena,ibenb}.
Giants having A(Li) exceeding this value are usually termed Li-rich and
represent a challenge for standard stellar evolution models.

About 1-2\,\% of K-giants are Li-rich, according to the above definition
\citep[][]{brown89,smith95}. In order to explain their existence, it was
suggested that dilution might not have been effective in these stars or that the
atmospheric Li abundance might have been increased by the ingestion of a planet
or a brown dwarf because of the expansion of the stellar atmosphere
\citep[][]{siess99}. Alternatively, a Li production phase following the first
dredge-up dilution has been proposed.

The possibility that Li-rich giants avoided dilution and simply preserved their
initial lithium abundance is contradicted, however, by the low $^{12}$C/$^{13}$C
carbon isotopic ratios measured \citep[][]{dasilva95}, which suggests that
mixing has taken place in those stars. Moreover the existence of a few stars
with Li abundances exceeding the meteoritic value \citep[``super'' Li-rich
giants, e.g.][]{balachandran00} argue against a preservation of the primordial
Li abundance.  At the same time, the ingestion of a planet or a brown dwarf
would also lead to an increase of the  $^9$Be abundance, which is not observed
in Li-rich giants \citep[][]{melo05}. 

On the other hand, there is agreement among the investigators that Li can be
enhanced in the giant's atmosphere through the \citet[][hereafter
CF71]{cameron71} $^7$Be­--transport mechanism. In order for this mechanism to be
effective, $^3$He should first be circulated from the stellar atmosphere down to
regions at temperatures high enough for $^3$He burning through the 
$^3$He($\alpha$, $\gamma$)$^7$Be reaction. Then, the produced $^7$Be should be
circulated up to the stellar surface where it can decay into $^7$Li by electron
capture: $^7$Be(e$^-$, $\nu$)$^7$Li. In luminous (M$_{bol}$=-6 to -7)
intermediate mass asymptotic giant branch (AGB) stars, the H-burning shell is in
contact with the stellar convective envelope and, therefore, $^3$He burning
takes place  partially under convective conditions (hot-bottom burning). In
low-mass stars, the convective envelope and the  H-burning shell are, however,
not in touch and the existence of a mixing mechanism has to be postulated to
connect material from those two zones.

Besides the standard mixing caused by the first dredge-up, an additional
extra-mixing episode is known to take place at the luminosity bump along the red
giant branch (RGB) in low-mass stars \citep[][]{gratton04}. \citet[][hereafter
CB00]{charbonnel00} took advantage of Hipparcos parallaxes to accurately place
Li-rich giants in the Hertzspung-Russell diagram, and revealed that some stars
were in the process of completing the dilution and were, hence,  erroneously
classified as Li-rich. Besides those stars, truly Li-rich giants tend to cluster
into two groups: at the RGB-bump luminosity in low-mass stars and at the early
AGB in intermediate mass stars. At these phases the mean molecular weight
discontinuity left over by the first dredge-up is erased, allowing the
occurrence of extra-mixing processes. Most of the Li-rich giants present,
however, $^{12}$C/$^{13}$C compatible with  first dredge-up predictions (CB00).
The Li-rich stage would then be produced by a relatively shallow extra-mixing
episode, a precursor of the extra-mixing phenomenon that further lowers the
$^{12}$C/$^{13}$C isotopic ratio in more evolved stars
\citep[see][CB00]{charbonnel98}. After the stellar atmosphere is enriched in
lithium, convection begins to again dilute it by exposing surface material to
the higher temperatures of the stellar interiors. Therefore, the Li-rich phase
would be by nature a short-lasting phase, which agrees with the low number of
detected stars. $^{12}$C/$^{13}$C isotopic ratios lower than first dredge-up
prediction may indicate that the dilution process has started again and Li
should be lower than the peak value.

\citet[][]{charbonnel00} certainly provide a valuable framework for the
investigation of Li-rich giants. Nevertheless, the authors do not embrace the
whole complex phenomenology of Li-rich giants. For instance, HD\,77361 is a
super Li-rich star lying close to the RGB-bump, but with $^{12}$C/$^{13}$C=4
\citep[][]{kumar09}, i.e. significantly lower than expected from first dredge-up
models. Li-rich low-mass giants were also detected along the upper RGB/AGB in
globular clusters \citep[][]{kraft99,smith99}, the Galactic Bulge
\citep[][]{uttenthaler07,gonzalez09} and dwarf spheroidal galaxies
\citep[dSph,][]{dominguez04,monaco08}. In particular, \citet[][]{dominguez04}
and \citet[][]{uttenthaler07} studied a sample of C stars in the Draco dSph and
in the Galactic Bulge, respectively. In these cases, the high Li abundance
detected in a few stars was attributed to enrichment during third dredge-up
episodes in thermally pulsating AGB stars. Notice also that Li-rich stars
detected in globular clusters and in the the tidal stream of the Sagittarius
dSph \citep[][]{monaco08} have low mass and lie either close to the tip of the
RGB or on the AGB. As such, they do not belong to the second category of Li-rich
giants identified by CB00. 

Rotation has often been claimed to be the culprit for extra-mixing processes
that happen along the RGB \citep[][]{sweigart79}. \citet[][]{fekel93} found that
many rapidly rotating ($v$\,sin\,$i$=6-46\,\kms), chromospherically active
giants present high Li abundances. They argued that the high observed rotational
velocities are the results of momentum transferred to the surface from the
stellar interior. This transfer may be accompanied by the dredge--­up of $^7$Be,
whose decay increases the surface Li abundance. Indeed, while K-giants are
usually slow rotators, \citet[][]{drake02} concluded that the fraction of
Li-rich giants may rise from the standard 1-2\% up to $\sim$50\% when rapid
rotators are considered ($v$\,sin\,$i$$>$8\,\kms).

Using IRAS colors, \citet[][]{delareza96,delareza97} also noted that many
Li-rich  stars presented far-infrared excess and speculated that the Li-rich
phase was associated with a mass-loss episode. A few Li-rich giants certainly
present evidence of mass loss and chromospheric activity
\citep[][]{balachandran00,drake02}. \citet[][]{fekel98} and
\citet[][]{jasniewicz99}, however, analyzed samples of giants with far-infrared
color excess and did not detect any additional Li-rich star. Therefore, mass
loss may or may not be associated with the Li-rich phase.

Several models were proposed to interpret the Li-rich giant phenomenon. 
While the classical Cameron-Fowler mechanism was envisaged to take place in AGB
stars (CF71) during He-shell flashes, the ``cool bottom processing" takes place
in RGB stars and can go on for an extended period of time as the star climbs up
the RGB \citep[][]{sackmann99}. In order to be able to produce $^7$Li, it
requires deep mixing with a high-speed $\dot{M}_p$, typically $\dot{M}_p \ga
10^{-7} M_\odot yr^{-1}$, but often as fast as $10^{-4} M_\odot yr^{-1}$.
According to the computations of \citet[][]{sackmann99}, the $^7$Li  in the
envelope increases as time on the RGB passes, reaches a plateau, and then
decreases again, as the freshly formed Li is mixed back into high-temperature
regions where it is destroyed. Thus the Li-rich phase is limited in time. The
level of Li-enrichment depends on the mixing speed as well as on the mixing
geometry, i.e. the relative surfaces of upward and downward streams.

According to \citet[][]{denissenkov00} and \citet[][]{denissenkov04}, enhanced
extra-mixing may be triggered by the episodic ingestion of a massive planet or a
brown dwarf \citep[see][]{carlberg10}. This could happen at any time along the
RGB. The ingested body would naturally contribute its momentum to the stellar
structure and, after the enhanced extra-mixing gave rise to the Li-rich
phenomenon, lithium would start to be destroyed again by convection. This model
would explain the existence of Li-rich giants all along the RGB at luminosities
different from the RGB-bump and the connection with stars with high rotational
velocities, but would neither explain the clustering of giants at the RGB-bump
in low-mass stars nor the existence of Li-rich giants with low rotational
velocities. 

Other possible sources for the extra-mixing episodes required to produce Li-rich
stars were identified in thermohaline mixing \citep[][]{charbonnel07} and
magneto-thermohaline mixing \citep[][]{denissenkov09}. \citet[][hereafter
GPBU09]{guandalini09} investigated the latter case and showed that both Li-rich
and Li-poor RGB/AGB stars can be accounted for by models of extra-mixing induced
by magnetic buoyancy. Their investigation does not try to interpret super
Li-rich giants (A(Li)$>$2.5\,dex) and still depends on several ad-hoc
assumptions, yet it provides a self-consistent interpretation for the apparent
clustering of Li-rich stars at the RGB-bump, as well as the presence of Li-poor
stars and the few Li-rich stars in the late RGB. The clump of Li-rich stars
detected on the early AGB by CB00 were interpreted by GPBU09 as stars that
actually are on the RGB. The first detection of a variable magnetic field in the
Li-rich giant HD\,232862 has been recently made \citep[][]{lebre09}.

Given the complexity of the problem, it is mandatory to collect more data to
better clarify the observational framework. Unfortunately, the intrinsic rarity
of Li-rich giants requires a dedicated search for these kind of objects, which
demands extensive observational campaigns.

In the context of a survey of the Galactic thick disk, we constructed a large
database of high-resolution, high-quality echelle spectra of giant stars
\citep[][]{carraro05,monibidin09,monibidin10}. By screening this database we
serendipitously detected the five Li-rich giants we present in the present work.
The paper is organized as follows: in \S\ref{sec:DataReduction} we outline the
target selection criterion and present the observations and data reduction
performed with reference to the detected Li-rich stars. In section
\S\ref{sec:AbundanceAnalysis} we present the performed abundance analysis on the
selected stars and in \S\ref{sec:RotVel} we present the measured target
rotational velocities. In section \S\ref{sec:ThickMembership} we use the derived
chemical abundances and atmospheric parameters together with their radial
velocities and proper motions to investigate whether the kinematics of our stars
are compatible with membership to the thick-disk. Finally, in sections 
\S\ref{sec:Discussion} and \S\ref{sec:Conclusions} we discuss and summarize our
findings.

\section{Observations and data reduction \label{sec:DataReduction}}

Our investigation is based on the sample defined by \citet[][hereafter
G06]{girard06}. It consists of 1196 red giant stars within 15 degrees from the
south Galactic Pole, with infrared 2\,MASS photometry \citep[][]{skrutskie06}
and absolute proper motions from the SPM3 catalog \citep[][]{girard04}.

The sample was defined from an IR color-magnitude diagram in the color range
0.7$\leq$J-K $\leq$1.1, to select intermediate metallicity stars, i.e.
thick-disk members. Contamination by halo stars should account for $\sim$8\% of
the selected objects (G06). The sample is confined to a cone perpendicular to
the Galactic disk, which G06 estimated to be volume-complete up to at least 3
kpc from the Galactic plane. A sloped-cutoff limit at faint magnitudes avoided
contamination by dwarf stars, except for the closest ones (d$\leq$63\,pc).

High-resolution, high-quality echelle spectra were acquired with four different
instruments for a total of 824 stars belonging to the G06 sample during 38
observing nights between 2005 and 2007. Telescopes with different collecting
powers were employed to efficiently secure spectra with roughly the same
resolution and quality for targets spanning a wide range of magnitudes
(5$\leq$V$\leq$16). 

We inspected by eye all spectra in the database in the region around the lithium
resonance doublet at 6707.8\,\AA. The line was detected in several stars, but
only the five objects in Table\,\ref{star} presented a strong line. Indeed, the
Li subordinate line at 6103.6\,\AA\, was detected in these stars only. A
detailed analysis of all objects with detected Li lines will be presented
elsewhere. We focus here on the five stars that alone present strong lithium
lines.

Stars discussed here were observed either with the Echelle/duPont or the
FEROS/2.2m spectrographs. Below we briefly summarize the data reduction
performed with reference to these observations. Details can be found in
\citet[][]{monibidin09} and \citet[][]{monibidin10}. In Table\,\ref{log} we
present the SPM identification number (which will be used henceforth), 2\,MASS
designations, date of the observations, instruments,  observatory, exposure
times, and resolutions for our stars. Table\,\ref{star} reports the target
coordinates, the 2\,MASS infrared magnitudes, the SPM3 proper motions and the
reddening as derived from the \citet[][]{sfd98} maps.

\input{log.tex}

\input{stars.tex}

During each run we observed several bright standard stars with accurate
parameters (radial and rotational velocities, fundamental parameters and
metallicity) available in the literature. Among them, the well known Li-rich
giant HD\,787 \citep[][]{castilho00} was also observed. This star is analyzed
here as well. 

Data reduction  was performed using standard IRAF\footnote{IRAF is distributed
by the National Optical Astronomy Observatories, which are operated by the
Association of Universities for Research in Astronomy, Inc., under cooperative
agreement with the National Science Foundation.} tasks and included  bias,
flat-fielding, order tracing, background and sky subtraction, and extraction.
Standard Th-Ar arcs were used for the wavelength calibrations. For the duPont
spectra we used ``milky'' flats, obtained with a diffusion filter set before the
CCD. These frames provided the needed  wavelength-independent pixel-to-pixel
response variation but, at odds with standard spectroscopic flat-field frames,
the ``milky'' flats did not correct the fringing pattern that affected our
spectra longward of about 7500\,\AA. For these, we created a fringing image from
the Dome Flat, reduced as a normal spectrum and then normalized, and we divided
all science spectra by this image in the reddest orders, where the fringing
pattern became visible. The resulting correction was deemed satisfactory. 

The radial velocities (RVs) of the observed stars were measured with a
cross-correlation (CC) technique \citep[][]{tonry79} by means of the fxcor IRAF
task. Radial velocity standards acquired with the same instrument were used as
templates for the CC. After the radial velocity measurements, all spectra were
shifted to the laboratory wavelength. The IRAF task {\tt rvcorrect} was used to
calculate the Earth motion and convert the observed RVs to the heliocentric
system. Measured RVs  are reported in Table\,\ref{PA} as well as the spectral
signal-to-noise (S/N).

\input{param.tex}

\section{Chemical abundance analysis}\label{sec:AbundanceAnalysis}

For each target we obtained an initial estimate of the stellar effective
temperature (\teff) from the dereddened infrared (J-K) color along with the
\citet[][]{alonso99} calibrations. The reddening was derived from the COBE
reddening maps \citep[][]{sfd98}, obtaining E($J-K$) from E($B-V$) by means of
the transformations of \citet{cardelli89}. The surface gravity (log\,g) was
derived with the aid of theoretical isochrones \citep[][]{marigo08}, adopting an
age of 10\,Gyr \citep[see also][]{feltzing09}, a metallicity Z=0.004, and a
distance estimated from a photometric parallax, by assuming that target stars
followed the same (J-K) {\em vs} K mean ridge line as the globular cluster
47\,Tuc. Indeed, the thick disk age and metallicity are similar to those of this
cluster \citep[][]{wyse05}. The microturbulent velocity $\xi$ was evaluated from
the derived gravity adopting the scaling relation: $\xi$ (\kms)= -0.254 $\times$
log\,g + 1.930 \citep[][hereafter M08]{marino08}. 

These initial parameters were adopted to compute a proper model atmosphere for
each star using the the Linux port of the ATLAS-9 code
\citep[][]{k93,sbordone04}. Equivalent widths (EWs) of a selected sample of iron
lines (see M08) were measured with a Gaussian fitting. For each star, the
measured EWs and the model atmosphere were then used within the
MOOG\footnote{\url{http://verdi.as.utexas.edu/moog.html}} \citep[][]{sneden73}
local thermodynamic equilibrium (LTE) stellar line analysis code to refine the
atmospheric parameters. In particular, the stellar \teff\, was fixed by imposing
that the abundance derived for each line should be independent of the transition
excitation potential. $\xi$ was estimated by minimizing the dependence of the
abundances from the measured EWs, and log\,g by requiring that  Fe\,I and Fe\,II
lines gave the same iron abundance, within the errors. The stellar atmospheric
parameters eventually adopted are reported in Table\,\ref{PA} together with the
derived iron abundance.

The abundance of Li, C, N, O, and the $^{12}$C/$^{13}$C isotopic ratio were
obtained with spectral synthesis of selected regions using the MOOG code along
with an ATLAS-9 model atmosphere computed for each star with the final adopted
atmospheric parameters. The atomic and molecular linelists were taken from the
Kurucz database\footnote{\url{http://kurucz.harvard.edu/}}. 

For the hottest stars in the sample, i.e. \#142173, \#313132, and \#343555, it
was possible to measure the carbon abundance from the C$_2$ band at 5635.5\,\AA.
For stars \#171877, \#225245 and HD\,787 carbon was measured from the CH G-band
at 4310\,\AA. For \#343555 we managed to measure carbon from both bands and
obtained an abundance difference of 0.05\,dex, which proves the consistency of
the two measurements. Oxygen was measured by synthesis of the line at 6300\,\AA.
The  Ni\,I line at 6300.34\,\AA\, was taken into account in the synthesis of the
Oxygen line. For the Ni\,I line we adopted the log\,gf value derived by
\citet[][]{johansson03}, namely log\,gf=-2.11. The nitrogen abundance was
derived by synthesis of the 8001-8005\,\AA\, spectral region, which is rich in
CN lines. The same region was also used to set the $^{12}$C/$^{13}$C isotopic
ratio, thanks to presence of the $^{13}$CN line at $\sim$8004.7\,\AA\, and the
$^{12}$CN lines at 8003-8004\,\AA. Owing to the low residual intensity of the 
$^{13}$CN line, for stars \#142173, \#171877, and \#343555 we provide only 
lower limits to the carbon isotopic ratio. 

Because the C-N-O elements are partly bound together in molecules for stars
cooler than $\sim$4500\,K, their abundances were derived together in an
iterative process. Solar abundances for Fe, C, N, and O were obtained in the
same way using the Kurucz solar spectrum for the purpose \citep[][]{k84}. The
C$_2$ band at 5635.5\,\AA\, is not visible in the solar spectrum, so we used the
CH G-band. The derived solar abundances are reported in Table~\ref{Ab}.

Afterward, we measured the lithium abundances from both the resonance doublet at
6707.8\,\AA\, and the subordinate line at 6103.6\,\AA. For the former, we
adopted the \citet[][]{reddy02} hyperfine structure. We also added to the
linelist the two Fe\,I and Ti\,I lines at 6707.433\,\AA\, and 6708.125\,\AA\,,
respectively, listed in \citet[][hereafter G09]{gonzalez09}. These lines are not
present in the Kurucz database, but are available at the Vienna Atomic Line
Database
\citep[VALD\footnote{\url{http://vald.astro.univie.ac.at/~vald/php/vald.php}},][]{vald}
and are required to obtain a good fit of the Li\,I doublet at 6707.8\,\AA. For
these lines we adopted the oscillator strength (log\,gf) calibrated on Arcturus
by G09. 

\begin{figure}
\includegraphics[width=1\columnwidth]{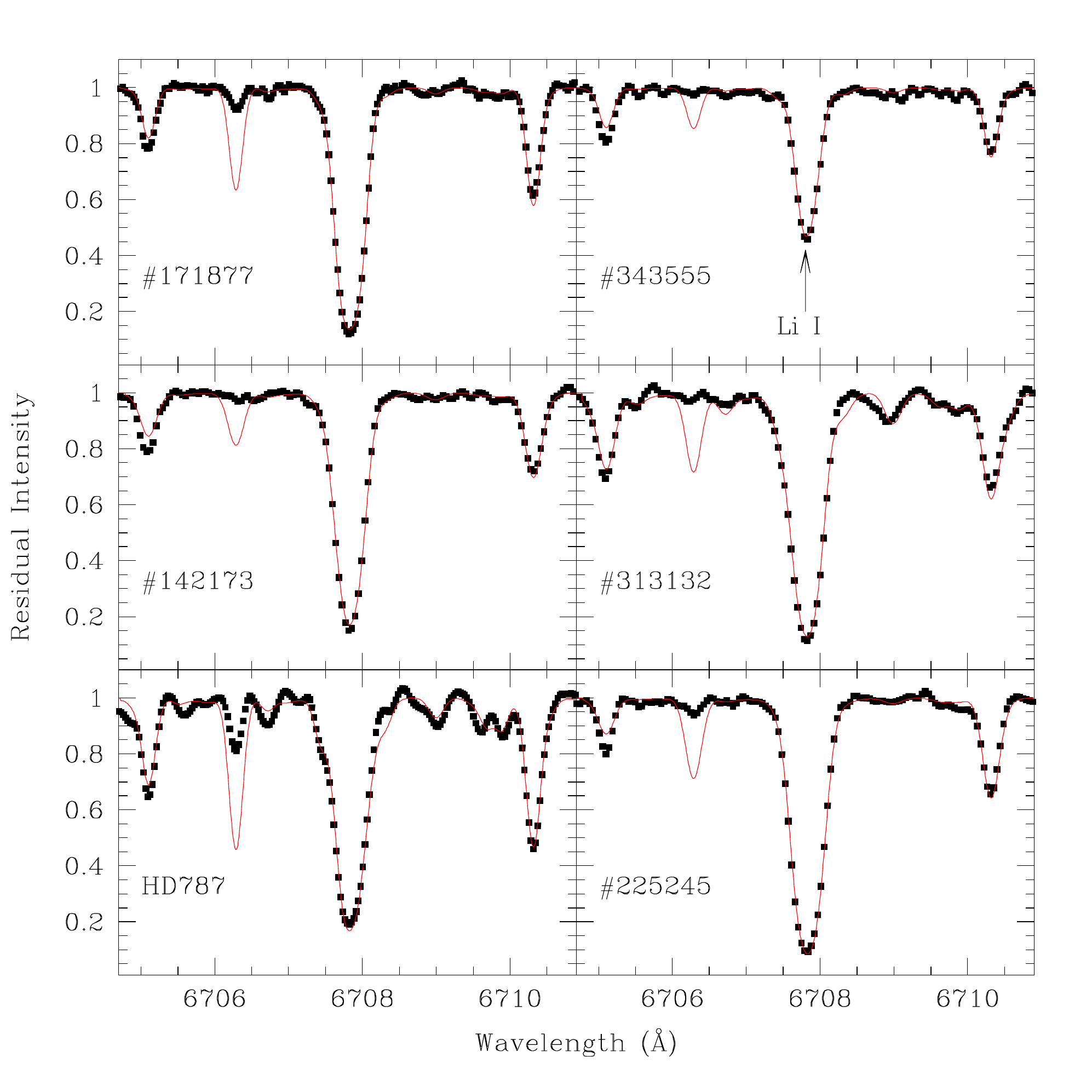}
\caption{Sample of the target stars' spectra in the region of the Li\,I resonance
doublet.}\label{li670}
\end{figure}

\begin{figure}
\includegraphics[width=1\columnwidth]{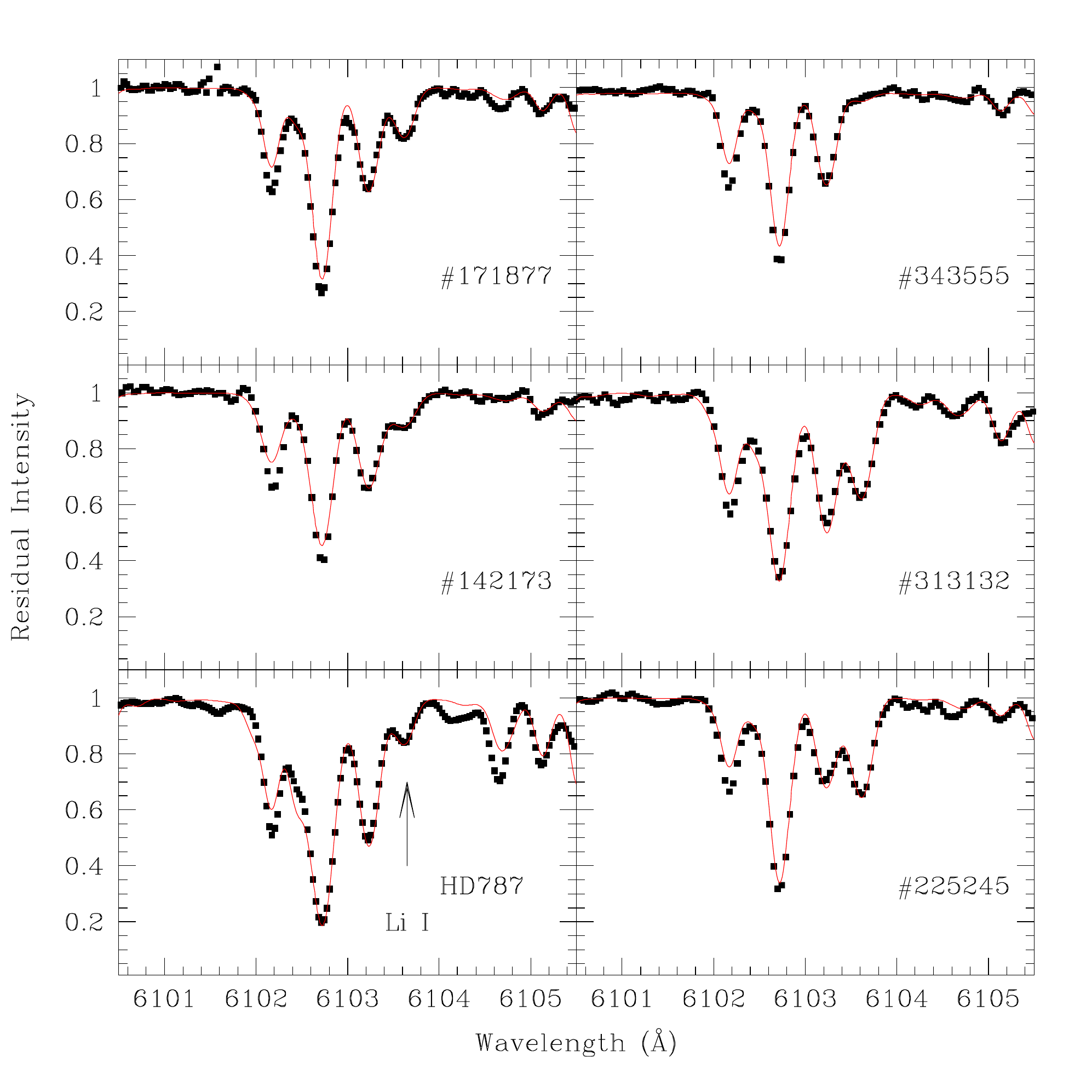}
\caption{Sample of the target stars' spectra in the region of the Li\,I
subordinate line.}\label{li610}
\end{figure}

\begin{figure}
\includegraphics[width=1\columnwidth]{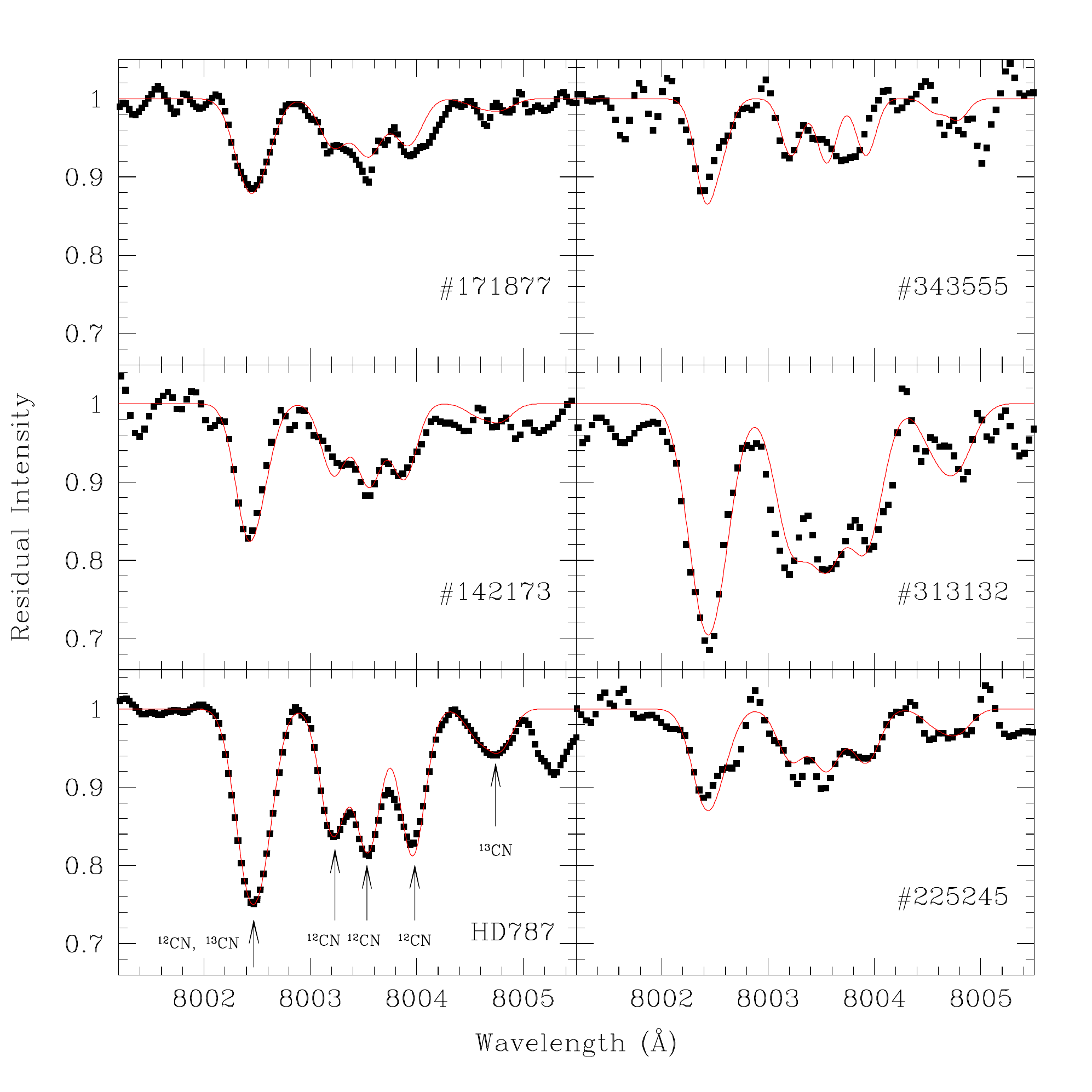}
\caption{Sample of the target stars' spectra in the region used to measure the N
abundance and the $^{12}$C/$^{13}$C isotopic ratio.}\label{c1213}
\end{figure}

Figures \ref{li670}, \ref{li610}, and \ref{c1213} present the spectra of our
stars and HD\,787 around the two Li\,I lines and in the 8001-8005\,\AA\,
region.  For each star, the best-fitting synthetic spectrum (continuous line) is
superimposed on the observed one (filled squares). The measured C, N, O, and Li
abundances are reported in Table\,\ref{Ab}. For star \#343555, the lithium
abundance was measured from the 6707.8\,\AA\, line, while we derive an upper
limit only from the subordinate line, which  is barely detected in the observed
spectrum.

\input{ab.tex}

The first of the two HD\,787 entries in Table\,\ref{Ab} reports the Li
abundances we obtained adopting the atmospheric parameters in \citet[][see first
HD\,787 entry in Table\,\ref{PA}]{castilho00}. We derived Li abundances of 2.03
and 2.14\,dex for the 6707.8\,\AA\, and 6103.6\,\AA\, lines, respectively. This
agrees well with \citet[][]{castilho00}, who measured 2.0 and 2.2\,dex (LTE
analysis). These authors did not measure the C, N, and O abundances and,
therefore, we adopted the \citet[][]{melendez08} values, which are also reported
in the Table. We also report the $^{12}$C/$^{13}$C isotopic ratio measured by
\citet[][]{dasilva95}. 

For HD\,787, however, we derive a slightly cooler temperature (see second
HD\,787 entry in Table\,\ref{PA}) with respect to \citet[][]{castilho00}. We
therefore performed an independent abundance analysis, following the same
procedure as outlined above. The second HD\,787 entry in Table\,\ref{Ab} reports
the C, N, O, and Li abundances we derived for HD\,787 with our analysis. The
$^{12}$C/$^{13}$C isotopic ratio we measured also perfectly agrees with
\citet[][]{dasilva95}.

\citet[][]{lind09} calculated the correction to the Li abundances measured from
both the resonance and the subordinate lines at 6707.8\,\AA\, and 6103.6\,\AA\,
due to non-LTE effects (NLTE). We used their grid to evaluate the correction to
apply to our measures. In columns 10 and 11 of Table\,\ref{Ab} we report the Li
abundances corrected for NLTE. These values have to be regarded only as
referential. Indeed, owing to the grid boundaries, we had to adopt in the
computation $\xi$=2.0\kms for all stars; \teff=4000\,K for \#171877,  \#225245,
and HD\,787 (2nd instance); log\,g=1.0 for \#225245 and HD\,787 (2nd instance);
[Fe/H]=0.0 for \#313132 and HD\,787 (2nd instance).

We finally estimated the errors on absolute abundances and abundance ratios
owing to the noise of the spectra and to the uncertainties on atmospheric
parameters.  For the $^{12}$C/$^{13}$C ratio only the error owing to the
spectral S/N is significant because a change in temperature, gravity, or
microturbulence does not alter the relative intensity of the $^{12}$CN or
$^{13}$CN molecular lines. For Fe, the uncertainty owing to the noise is simply
the error on the mean as given by MOOG. For the other elements, whose abundances
were obtained by spectrosynthesis, the errors were evaluated by comparing
observed spectra with synthetic ones calculated with a step of 0.05 dex in
abundance or 2 in $^{12}$C/$^{13}$C.  We estimated by eye the abundance interval
that is still compatible, within the noise, with the observed spectrum. The
error owing to the noise was assumed to be the semi-amplitude of this interval.

As for the atmospheric parameters, uncertainties were estimated by changing one
parameter at the time and measuring the abundances again. The error for a given
parameter is given by the difference between the old and the new abundances.  We
assumed $\pm$100 K, $\pm$0.2 dex, and $\pm$0.10 km/s as the typical errors for
\teff, log\,g and $\xi$, respectively, and analyzed stars \#313132 and \#225245,
which cover the entire \teff\, range of our sample. Results are reported in
Table\,\ref{ER}. The last column in the table reports the total error, which is
defined as

\begin{center}
$\sigma_\mathrm{tot}=\sqrt{\sigma_\mathrm{S/N}^2+\sigma_{T_\mathrm{eff}}^2+\sigma_\mathrm{log(g)}^2+\sigma_\mathrm{\xi}^2}$.
\end{center}

\input{errors.tex}

\section{Rotational velocities}\label{sec:RotVel}

The projected rotational velocity of the target stars, $v$\,sin\,$i$, was
measured from the width of the auto-cross correlation function (CCF) of their
spectrum: each one was cross-correlated \citep{tonry79} with itself through the
{\it fxcor} IRAF task, and the peak of the resulting function was fitted with a
Gaussian profile to derive its width $\sigma$. We verified that the resulting
width changed when correlating different spectral ranges, but only by the
addition of a color-dependent constant on all standard and target stars. This
did not affect the measurements, since it only implied a different value for the
calibrated constants of the $\sigma$-$v$\,sin\,$i$ relation. We finally adopted
the range 6000-6500 \AA, a high-S/N region free of telluric bands and strong
stellar lines, avoided because of their non-Gaussian wings.

The square of the width of the auto-CCF ($\sigma_\mathrm{obs}$) can be expressed
as the quadratic sum of the rotational broadening ($\sigma_\mathrm{rot}$) and a
term including all the causes of line broadening other than rotation
($\sigma_0$). The relation between the measured $\sigma_\mathrm{obs}$ and the
projected rotational velocity can thus be written as

\begin{equation}
\sigma_\mathrm{rot}^2=k\cdot(v\sin{i})^2=\sigma_\mathrm{obs}^2-\sigma_0^2,
\label{evsini1}
\end{equation}

\noindent where $k$ is a constant coupling $v$\,sin\,$i$ to the line rotational
broadening. $\sigma_0$ is the sum of an instrumental constant and another term
depending on the physical conditions in the stellar atmosphere (temperature,
gravity, macro- and micro-turbulence). As a first approximation, $\sigma_0$ can
be expressed as a function of color \citep{Benz84,Queloz98}. Previous
investigations showed that a linear relation between $\sigma_0$ and $(J-K)^2$ is
the most appropriate \citep{Melo01} but, having at our disposal only five red
giant standard observed for the calibration of the coefficients, we preferred
the lower-order expression $\sigma_0^2=a+b\cdot(J-K)^2$. This approximation is
allowed because our target stars are distributed in a narrow range of colors,
and the systematics caused by a lower-order fit are expected to be negligible.
In summary, renaming the involved constants appropriately, the relation between
the measured $\sigma_\mathrm{obs}$ and $v$\,sin\,$i$ can be written as

\begin{equation}
(v\sin{i})^2=A\cdot\sigma_\mathrm{obs}^2+B+C\cdot(J-K)^2.
\label{evsini2}
\end{equation}

The coefficients (A,B,C) were calibrated through a two-variables least-square
procedure, by means of the five red giants standard observed during the same
runs as our targets, with $(J-K)$=0.50-0.85 and projected rotational velocity
from the catalog of \citet{Glebocki00}. The FEROS and duPont runs were
calibrated separately, because of their different instrumental broadening.

The validity of the solution was tested comparing the $v$\,sin\,$i$ of the
standard stars calculated through Equation~\ref{evsini2} with its literature
value. The results are given in Table~\ref{tvsinistand}. The mean difference is
negligible and the dispersion of the differences with literature values is 0.6
km~s$^{-1}$.

The rotational velocities measured for our program stars are given in
Table~\ref{Ab}. The errors were estimated varying $(J-K)$,
$\sigma_\mathrm{obs}$, and the input quantities of the calibration (color,
$v$\,sin\,$i$, and CCF width of the standard stars), by $\pm1\sigma$. The error
on $(J-K)$ was taken from the 2MASS catalog, while an uncertainty of 0.05
km~s$^{-1}$ was assumed for the measured $\sigma$, according to Equation~6 of
\citet{Melo01}. For very slow rotators, $\sigma_\mathrm{obs}\sim\sigma_0$ and
measurement errors can cause $\sigma_\mathrm{obs}<\sigma_0$, thus leading to an
unphysical negative rotational velocity. This happened for two target stars. The
upper limit of their $v$\,sin\,$i$ was estimated as
$v\sin{i}_\mathrm{max}=A\sqrt{2\sigma_0 \epsilon}$ \citep{Melo01},
where $\epsilon$ is the uncertainty on $\sigma_0$, evaluated with the same
procedure used to derive the errors on rotational velocity.

The derived projected rotational velocity is very low for all stars. This
general result is confirmed by a simple inspection of the CCFs, because the
$\sigma_\mathrm{obs}$ of the targets never exceeded the analogous quantity for
more than a couple of standard stars. Although the width also depends on the
spectral type, this indicates that $v$\,sin\,$i$ of all the stars must be in the
range spanned by the standards, i.e. 0-5 km~s$^{-1}$. The resulting errors are
very similar to the dispersion of the differences with the literature value for
the standard stars.

\begin{table*}
\begin{center}

\caption{Projected rotational velocity derived for the standard stars. The
literature $v$\,sin\,$i$, and the derived values for the FEROS and DuPont runs, are
given in columns 2-4. In the last column the literature references are given: 1:
\citet{Delareza95}; 2: \citet{demedeiros99}; 3: \citet{Demedeiros96}; 4:
\citet{fekel93}; 5: \citet{Fekel97}; 6: \citet{Gray82}; 7: \citet{Pasquini00};
8: \citet{Smith79}; 9: \citet{Gray86}. When more than one estimate was
available, the value in column~2 is their weighted average, with their standard
deviation as associated error.}

\label{tvsinistand}
\begin{tabular}{l| c c c | c}
\hline
\hline
ID & $v$\,sin\,$i$(lit.) & $v$\,sin\,$i$(FEROS) &  $v$\,sin\,$i$(duPont) & ref. \\
 & km s$^{-1}$ & km s$^{-1}$  & \\
\hline
HD\,787$^a$     & 2.1$\pm$1.0 & 1.3 & 1.5  & 1,2,3,4 \\
           	& 2.1$\pm$1.0 &  -  & 1.6  & \\
HD\,31767  	& 1.1$\pm$1.3 & 2.0 & 2.3  & 2 \\
HD\,180540 	& 4.8$\pm$0.3 & 4.7 & 4.8  & 9 \\
HD\,4128   	& 3.5$\pm$0.4 & 3.6 & 3.4  & 5,6,7,8 \\
HD\,15453  	& 1.0$\pm$1.0 & 0.9 & 1.3  & 2 \\
\hline
\end{tabular}
\end{center}

$^a$ For HD\,787 only the FEROS spectrum obtained in the run reported in
Table\,\ref{log} was used for the abundance analysis. However, the star was
repeatedly observed during the various duPont and La\,Silla runs. We report here
the rotational velocities measured for the different spectra.

\end{table*}

\section{Thick-disk membership}\label{sec:ThickMembership}

\input{distance_new.tex}

The stars under analysis are drawn from a sample of likely thick-disk stars, as
from the target selection criterion (see \S\ref{sec:DataReduction}). Further
insights into the actual nature of stars in our sub-sample can now be obtained
by coupling the star kinematics to stellar isochrones and the estimated
metallicities and atmospheric parameters.

We estimated the star distances as follows. For each star, we first generated an
appropriate isochrone\footnote{http://stev.oapd.inaf.it/cgi-bin/cmd}
\citep[][]{marigo08} using the metallicities in Table\,\ref{Ab}. An age of 
10\,Gyr was adopted for all stars at this stage \citep[][]{feltzing09}. Then, we
inspected the corresponding isochrone searching for the best match of both
gravity and effective temperature. This provides an estimate of the absolute 
magnitude which, together with the apparent magnitude and reddening in
Table\,\ref{star}, allows to derive the star distances. Allowing for an
uncertainty in the age of $\pm$2\,Gyr results in a typical error in distance of
about 10$\%$. The derived distances are listed in Table\,\ref{dist}. For further
reference, notice that a variation of $\Delta$\teff=$\pm$100\,K and
$\Delta$log\,g=$\pm$0.5 would result in a variation on the derived distances of
about 4\% and 17\%, in the case of \#142173 taken as representative of the stars
under investigation.

To test the consistency of the possible thick-disk membership of the target
stars, we explored their kinematical and orbital properties. Using the star
distances in Table\,\ref{dist} and the star radial velocities (Table\,\ref{PA})
and proper motions (Table\,\ref{star}), we  estimated the Cartesian Galactic
coordinates and velocity components (see Table\,\ref{dist}) following the
method  described by \citet[][]{johnson87}. Finally, we calculated the Galactic
orbit of each star adopting a gravitational potential including the bar,
deriving estimates for the eccentricity (e), the apo- and peri-galacticon
(R$_{a}$ and R$_{p}$), and the absolute value of the maximum height above the
plane Z$_{max}$, which are also indicated in Table\,\ref{dist}. Orbits were 
computed following the method described in \citet[][]{magrini10} and
\citet[][]{jilkova10}, which the reader is referred to for details.

Based on the above analysis, we can draw the following conclusions:

\begin{itemize}
\item according to the derived orbital parameters, all stars but \#313132 
look like typical thick-disk stars;
\item \#313132 has a low eccentricity and does not reach high Z. Besides, its
orbit is confined within the solar vicinity. It is, therefore, reasonable to
consider it as a thin-disk star. We note also that \#313132 presents a solar 
metallicity (see Table\,\ref{Ab}), which is typical of thin-disk stars.
\end{itemize}

\section{Discussion}\label{sec:Discussion}

The abundances measured for the target stars are summarized in Table\,\ref{Ab}.
All five stars have lithium abundances higher than 1.5\,dex, and are, therefore,
to be considered as Li-rich. Three of them present abundances exceeding 2.2\,dex
and one, namely \#313132, belongs to the group of the rare super Li-rich giants
\citep[see, e.g.,][]{balachandran00,dominguez04,reddy05,monaco08,kumar09}.

\begin{figure}
\includegraphics[width=1\columnwidth]{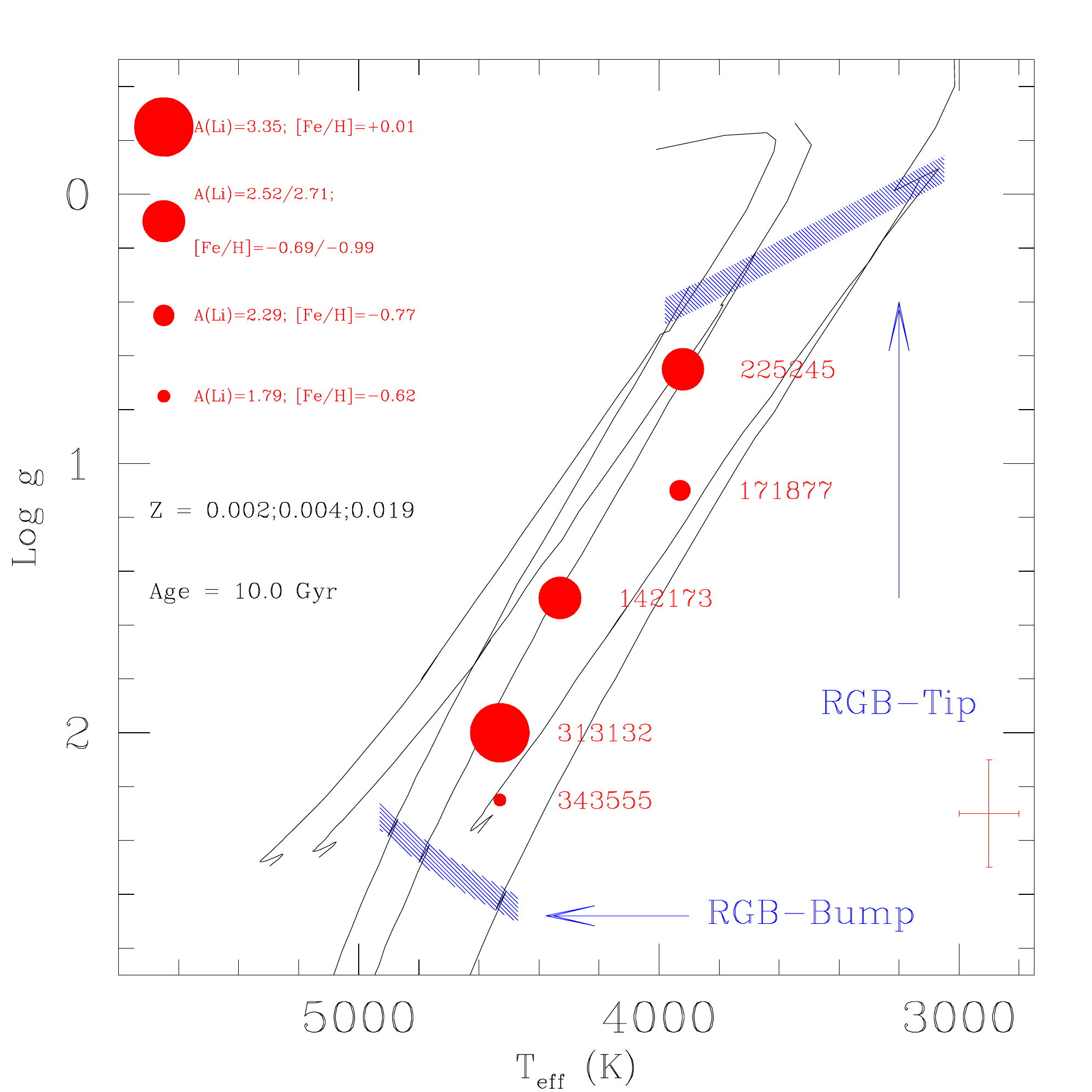}

\caption{Target stars are plotted in the log\,g {\em vs} \teff\, plane together
with 10\,Gyr theoretical isochrones of metallicity similar to the target stars.
The symbol size is scaled according to the stars' Li content.}\label{cmd}

\end{figure}
 
In Fig.\,\ref{cmd} we plot our stars in the \teff\, {\em vs} log\,g plane,
together with 10\,Gyr theoretical isochrones \citep[][]{marigo08} spanning the
metallicity range covered by our stars. Though they likely cover a range of
ages, \citet[][]{feltzing09} concluded that thick-disk stars are older than the
Sun with a mean age of 10\,Gyr. The position of the RGB-bump and RGB-tip are
marked in the figure for reference. For younger ages the RGB-bump position on
the plane moves to cooler temperatures and lower gravities, but  for ages as
young as 3\,Gyr, its location lies still within the shaded area in the figure. 

As discussed in section \S\ref{sec:ThickMembership}, \#313132 is likely to be a
thin-disk star, and we will discuss its case separately. The other four stars
seem to be more evolved than the RGB-bump, but lie below the RGB-tip.  
Their position in Fig.\,\ref{cmd} would suggest they might have already suffered
the extra-mixing process known to take place at the RGB-bump. While for stars 
\#142173, \#171877 and \#343555 we can only derive lower limits to the
$^{12}$C/$^{13}$C isotopic ratio, star \#225245 indeed presents a value lower
than first dredge-up prediction for low-mass stars. If the Li-enrichment phase
indeed occurs at the RGB-bump (CB00), this star may be in the process of
diluting again their surface Li abundance and would be, therefore, similar to
the HD\,148293 ($^{12}$C/$^{13}$C=16, A(Li)=2) and HD\,183492 cases
($^{12}$C/$^{13}$C=9, A(Li)=2) discussed by CB00. Following CB00, if it is
indeed diluting its surface Li, its position in Fig.\,\ref{cmd} and its carbon
isotopic ratios might provide constraints on the time-scale for the extra-mixing
process to occur. At variance with HD\,148293 and HD\,183492, however, \#225245
is significantly more evolved than the RGB-bump and have high lithium abundances
(A(Li)$>$2.7). 

Cool-bottom processing (CBP) could still be considered as a viable mechanism to
produce the lithium abundances observed in these four stars. In particular,
\#225245 has quite a high lithium abundance (A(Li)$>$2.7). It lies close to the
RGB-tip and has quite a low carbon isotopic ratio ($^{12}$C/$^{13}$C=8), similar
to the predictions made by \citet[][]{boothroyd99} for low-mass stars at the
RGB-tip using a model of deep mixing and the associated CBP. These models were
proved to be effective in producing Li-rich stars all along the RGB
\citep[][]{sackmann99} up to very high abundances, A(Li)=4.0. Considering its
low metal content, \#225245 may also be similar to the Li-rich giants discovered
in the globular clusters M\,3 and NGC\,362 \citep[][]{kraft99,smith99}.

\begin{figure}
\includegraphics[width=1\columnwidth]{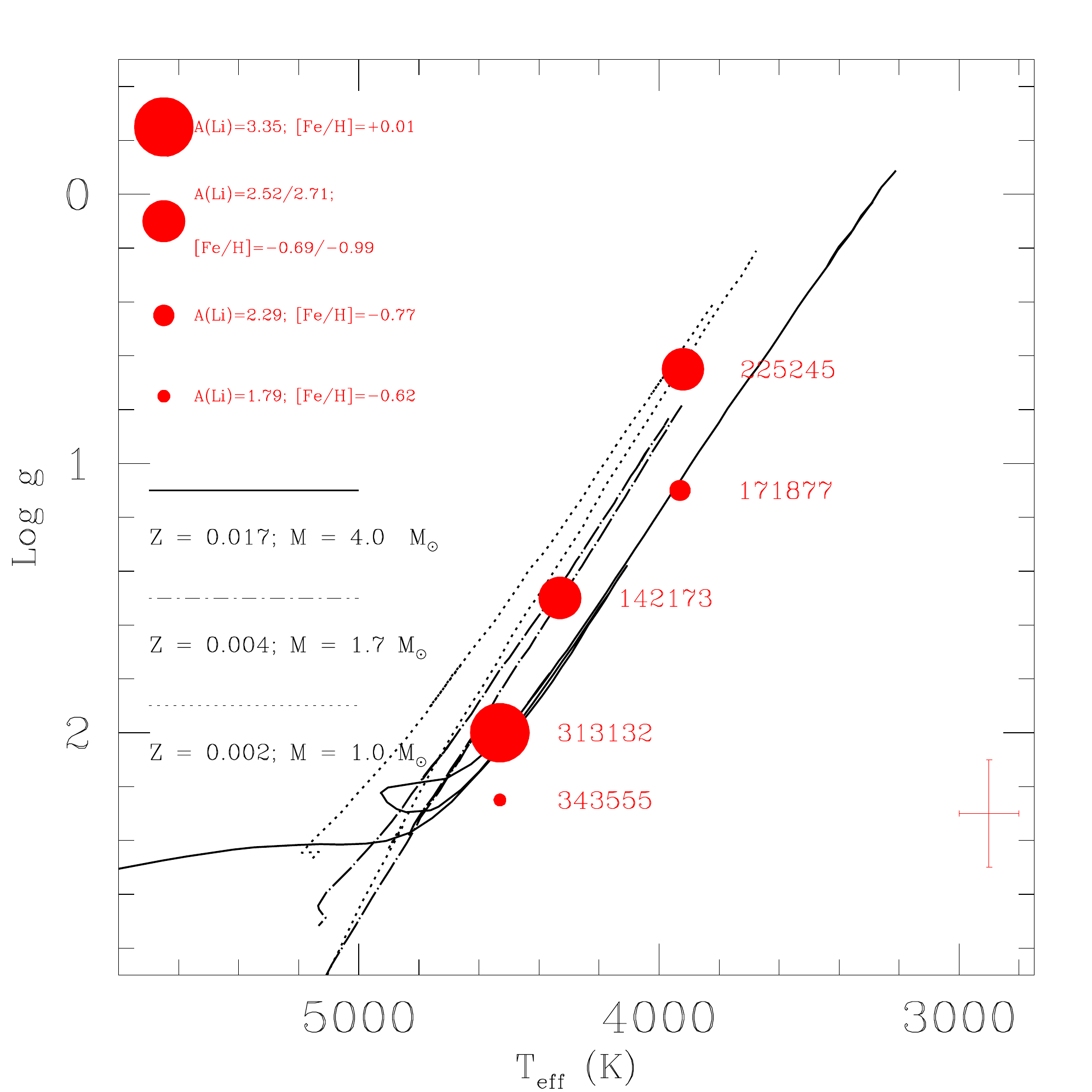}

\caption{Target stars are plotted in the log\,g {\em vs} \teff\, plane together
with theoretical evolutionary tracks of metallicity similar to the target stars
and different masses. The symbol size is scaled according to the stars' Li
content.}\label{cmd2}

\end{figure}

Together with our stars, we plot in Fig.\,\ref{cmd2} stellar evolutionary tracks
\citep[][]{bertelli08,bertelli09} of metallicity appropriate to our stars. 
While the arguments in section \S\ref{sec:ThickMembership} cannot be considered
as proof that they are thick-disk members, the locus occupied in the plane by
the four stars we just discussed is consistent with them indeed being low-mass
objects and, therefore, with the conclusions we have drawn above. We note that
CB00 has shown that Li-rich giants cluster either at the RGB-bump or at the
early AGB, in the case of low and intermediate-mass stars, respectively. Our
stars, however, do not belong to either of these two categories.

It is clear from Fig.\,\ref{cmd2} that the atmospheric parameters of star
\#313132 are perfectly compatible with a 4\,M$_\odot$ star. Its location in the
plane is compatible with either being on the RGB or the AGB. This object may
accordingly  belong to the second group of Li-rich stars identified by CB00,
i.e. stars on the early AGB, where the deepening of the envelope has not yet
reached its maximum penetration and an extra-mixing process is supposed to occur
to drive $^3$He from the envelope to a region hot enough for the
\citet[][]{cameron71} mechanism to take place. In this case, it would be similar
to HD\,787 and HD\,30834, which also have similar $^{12}$C/$^{13}$C isotopic
ratios, compatible with first dredge-up predictions for similar masses (see
discussion in CB00). Star \#313132 is, then, the first super Li-rich giant
belonging to this group. Indeed, stars on the early AGB present on average lower
Li abundances than stars at the RGB-bump. \citet[][]{guandalini09} have
interpreted this group of stars as actually lying on the RGB. Their models are
unable to interpret super Li-rich stars though. We notice that it is in the
4-8\,M$_\odot$ mass range that hot-bottom burning leading to Li production
starts to became effective. This is known to occur at bolometric magnitudes
M$_{bol}\simeq$ -6 to -7 \citep[][]{sackmann92,smith95}. Theoretical tracks and
isochrones \citep[][]{bertelli08,bertelli09,marigo08} suggest instead that 
atmospheric parameters similar to that of star \#313132 are compatible with 
significantly fainter luminosities only, namely M$_{bol}\simeq$-2.1 to -1.4, for
the following combination of parameters:
(\teff;log\,g;mass)=(4400-4600\,K;1.8-2.2;3.5-4.5\,$M_\odot$). Therefore, the
high lithium abundance of \#313132 should not have been reached through
hot-bottom burning.

All our stars but \#343555 have Li abundances that exceed or are close to the
maximum value allowed by the GPBU09 model for stars at the RGB-bump
(A(Li)$\simeq$2.5\,dex), while more evolved stars would present even lower
values.

As discussed in section \S\ref{sec:Introduction}, the fraction of Li-rich giants
seems to be highly enhanced among rapid rotators \citep[][]{drake02}. The
rotational velocities measured for our stars are reported in Table\,\ref{Ab}.
All our targets are slow rotators, with only the super Li-rich star \#313132
presenting a slightly higher value, namely 3.3\,\kms. This value is, however,
still well below the range of rotational velocities ($v$\,sin\,$i>$8\,\kms)
among which the frequency of Li-rich giants seems to increase.

\begin{figure}
\includegraphics[width=1\columnwidth]{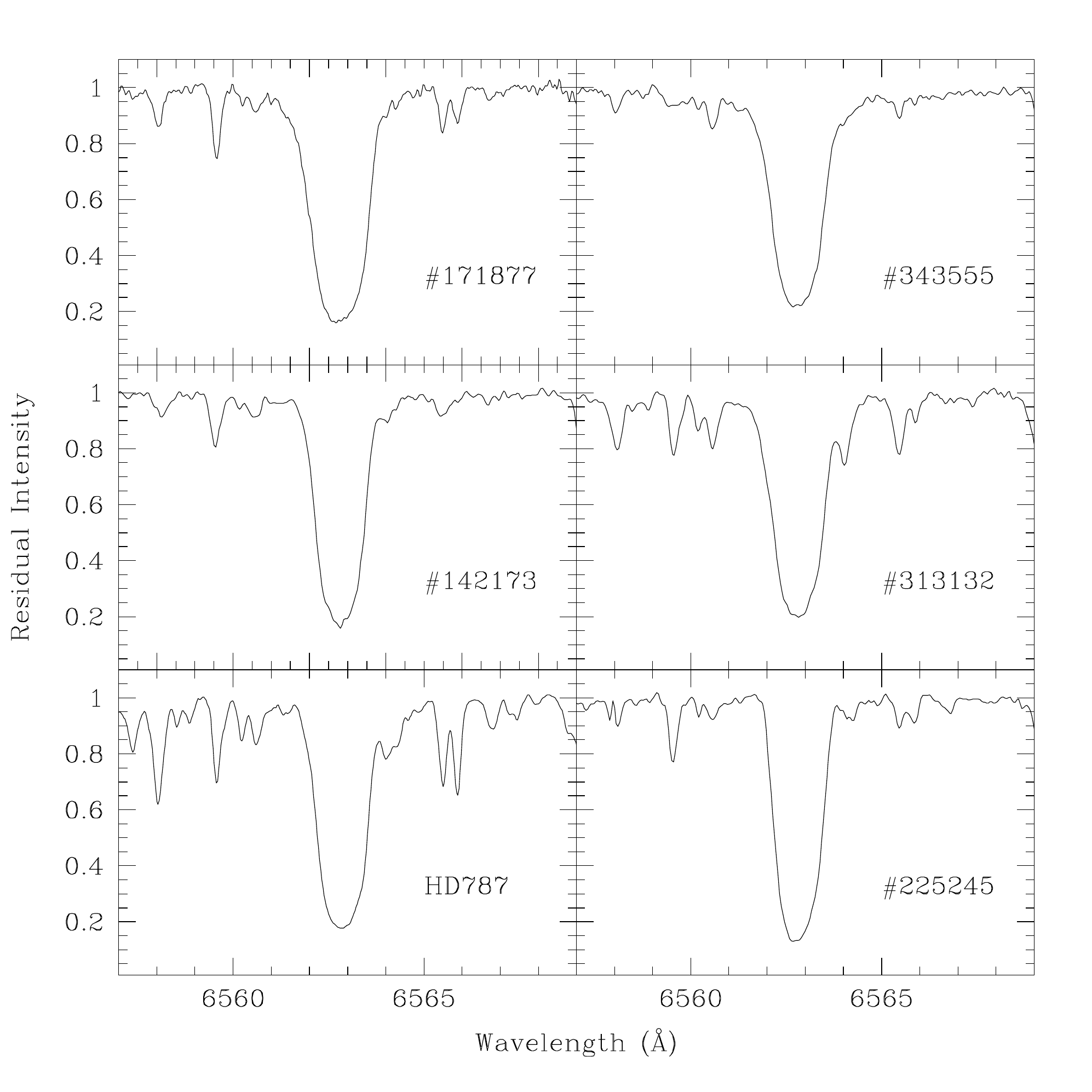}
\caption{Sample of the target stars' spectra in the region of the H$_\alpha$ 
line.}\label{ha}
\end{figure}

\begin{figure}
\includegraphics[width=1\columnwidth]{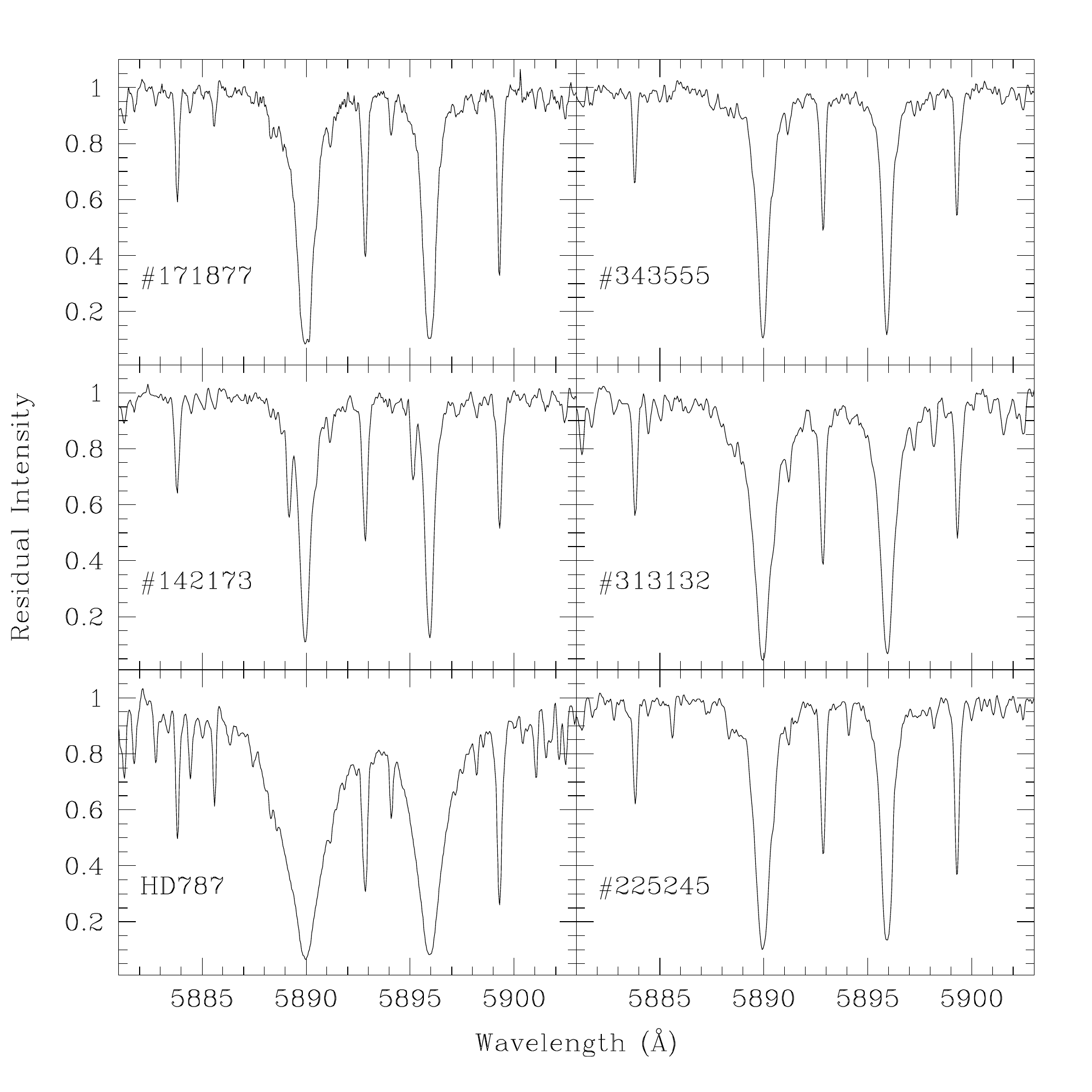}
\caption{Sample of the target stars' spectra in the region of the Na\,D
lines.}\label{na}
\end{figure}

In RGB stars, the Li-rich phase has also been suggested to be associated with a
mass loss-episode \citep[][]{delareza96,delareza97}. This latter can cause
blue-shifted asymmetric H$_\alpha$ profiles or the presence of additional
blue-shifted absorption lines in the Na doublet (Na\,D) lines
\citep[][]{balachandran00,drake02,reddy05} in the stellar spectra.
Figs.\,\ref{ha} and \ref{na} present a sample of the spectra obtained for the
target stars in the H$_\alpha$ and Na\,D lines regions. Our stars do not present
any evident asymmetry in the H$_\alpha$ line (Fig.\,\ref{ha}). \#142173 is the
only star that presents additional blue shifted lines around the sodium doublet
(Fig.\,\ref{na}). These additional components are blue-shifted by 40 and
39\,\kms with respect to the companion photospheric Na\,D lines. These
velocities closely correspond to the local standard of rest
($v_\mathrm{helio}$=+38\,\kms). Therefore, these features may originate in the
interstellar medium, and not be of stellar origin \citep[see,
e.g.,][]{balachandran00}.

\section{Conclusions}\label{sec:Conclusions}

We have built up a large database of high-resolution, high-quality stellar
spectra  of candidate thick-disk giants.  We searched it for stars presenting a
strong Li\,I resonance line and detected five Li-rich giants. One of them,
namely \#313132, has A(Li)$>$3.3 and therefore falls in the category of the rare
super Li-rich giants
\citep[see][]{kraft99,dominguez04,monaco08,kumar09,carlberg10}.

All our stars but \#313132 have kinematics compatible with thick-disk
membership, and a comparison with theoretical evolutionary tracks (see
Fig.\,\ref{cmd2}) of a proper metallicity is consistent with them being 
low-mass stars. They all seem to have evolved past the RGB-bump
(Fig.\,\ref{cmd}). Therefore, they do not belong to either of the two
evolutionary regions where Li-rich giants tend to cluster, namely the RGB-bump
or the early AGB in low- and intermediate-mass stars, respectively.
Fig.\,\ref{cmd} would suggest that these stars might have already undergone the
extra-mixing process known to happen at the RGB-bump. The low carbon isotopic
ratios measured in \#225245 may support this possibility, at least for this
star. Under this hypothesis, they can be either produced by CBP or are in the
process of diluting Li again after the enrichment phase. In the latter case,
their $^{12}$C/$^{13}$C isotopic ratios could be used as a diagnostic for the
time scale involved in the dilution process after the Li-enrichment phase has
occurred. \#225245 has a quite high lithium abundance (A(Li)$>$2.7) and lies
close to the RGB-tip. Its low $^{12}$C/$^{13}$C isotopic ratio is consistent
with the CBP predictions of \citet[][]{boothroyd99}.

\#313132 is a likely thin-disk member. Its spectroscopically derived atmospheric
parameters are compatible with a $\sim$4\,M$_\odot$ star either in the RGB or
AGB phase. It is the first super Li-rich giant detected at this phase (CB00,
GPBU09). Its $^{12}$C/$^{13}$C ratio is compatible with the first dredge-up
predictions for similar masses. 

The detected Li-rich stars are slow rotators and do not show indication of mass
loss, as shown by the inspection of the  H$_\alpha$ and the Na\,D lines. 

\begin{acknowledgements}  

C.M.B., S.V., and D.G. gratefully acknowledge support from  the Chilean {\sl
Centro de Astrof\'\i sica} FONDAP No. 15010003 and the Chilean Centro de
Excelencia en Astrof\'\i sica y Tecnolog\'\i as Afines (CATA). M.Z. acknowledges
Fondecyt Regular 1085278, the FONDAP Center for Astrophysics 15010003, Basal
CATA PFB-06, and the MIDEPLAN Milky Way Millennium Nucleus P07-021-F. We thank
Bruno Jungwiert for his input on a former version of the orbit integration code
used in this contribution. Finally, we would like to thank the referee, Dr. P.
North, for his helpful suggestions, which improved the quality of the present
investigation.

\end{acknowledgements}

\end{document}

%% file: log.tex
%\begin{sidewaystable}

\begin{table*}

\caption{Log of the observations for the program stars. The SPM and 2\,MASS
identification numbers are also reported together with the spectral resolution
of the obtained observations.}\label{log}

\begin{center}
\begin{tabular}{lcccclcccrr}
\hline
SPM  & 2MASS & Date & Exposure Time & $\lambda$/$\Delta \lambda$ & Instrument & Telescope & Observatory \\
&&& (s) &&&&&  \\
 \hline
\object{142173}        	    &  00321256-3834022   & 2006-09-13 & 1800 & 38\,000 & Echelle &  2.5m Ir\'en\'ee du Pont  & Las Campanas  \\
\object{171877}        	    &  00392022-3131354   & 2006-08-23 &  900 & 48\,000 & FEROS   &  2.2m  ESO-MPI	      & La Silla      \\
\object{225245}        	    &  00544638-2735304   & 2006-09-11 & 2400 & 38\,000 & Echelle &  2.5m Ir\'en\'ee du Pont  & Las Campanas  \\
\object{313132}$^a$ 	    &  01202066-3409541   & 2006-09-12 &  300 & 38\,000 & Echelle &  2.5m Ir\'en\'ee du Pont  & Las Campanas  \\
\object{343555}$^{b}$	    &  01294200-3015464   & 2006-09-12 &  850 & 38\,000 & Echelle &  2.5m Ir\'en\'ee du Pont  & Las Campanas  \\
\hline
\object{HD\,787}$^{c}$ 	    &  00120998-1756177   & 2006-08-20 &  100 & 48\,000 & FEROS   &  2.2m  ESO-MPI	      & La Silla       \\
\hline
\end{tabular}
\\
\smallskip
\end{center}
$^a$ Alternative ID: \object{CD-34 510}.	\\
$^b$ High-velocity star.\\
%$^c$ Alternative ID: \object{2MASS J01294200-3015464}.\\
%$^d$ Stellar photometry is obtained from the Simbad
%database\footnote{\url{http://simbad.u-strasbg.fr/simbad/}}. Reddening values
%are derived from the \citet[][]{sfd98} reddening maps.\\
$^c$ Alternative ID: \object{HR37}.
\\
\smallskip
\end{table*}				  
% HD787_LCO2_2scdr.fits	2006-09-11             	19.990  	        	             	
% HD787_LCO2scdr.fits  	2006-09-11             	40.010  	        	             	
% HD787_MIKEcdr.fits   	                       	5.000   	MIKE-Red	Clay (Mag.II)	
% s191965scdr.fits     	2005-09-22T  	        169.9999	FEROS   	MPI-2.2      	
% s338943cdr.fits      	                       	1300.000	MIKE-Red	Clay (Mag.II)	

%% file: stars.tex
%\begin{sidewaystable}

\begin{table*}

\caption{Coordinates, photometry, reddening and proper motions (pm) of the program stars.}

\label{star}
\begin{center}
\begin{tabular}{lccccccccrrrcrr}
\hline
SPM  & $\alpha$(J2000) & $\delta$(J2000)&    l	 &     b  & K & J-K & E(B-V)   & E(J-K) & pm(RA) & pm(Dec) &\\
&&				       & deg	 & deg    		       &&&&&mas/yr& mas/yr \\
 \hline
\object{142173}        		& 00:32:12.56 & -38:34:02.3& 321.099 & -77.874 &  9.29& 0.78 &  0.018& 0.009 & -1.2$\pm$2.0 & -19.4$\pm$1.6\\
\object{171877}        		& 00:39:20.23 & -31:31:35.5& 333.144 & -84.872 &  8.12& 0.91 &  0.019& 0.010 &  6.8$\pm$1.6 &  -7.7$\pm$2.5\\
\object{225245}        		& 00:54:46.38 & -27:35:30.4& 245.156 & -89.126 &  8.79& 0.88 &  0.014& 0.007 &  4.5$\pm$1.2 &  -3.3$\pm$1.1\\
\object{313132}         	& 01:20:20.66 & -34:09:54.1& 263.253 & -80.616 &  7.04& 0.73 &  0.031& 0.016 & 26.4$\pm$3.1 &  -5.2$\pm$2.4\\
\object{343555}         	& 01:29:42.01 & -30:15:46.4& 235.664 & -81.047 &  8.46& 0.72 &  0.017& 0.009 & 27.9$\pm$1.7 & -25.3$\pm$1.3\\
\hline
\object{HD\,787} 	    	& 00:12:09.99 & -17:56:17.8&  76.321 & -77.101 &  1.85& 0.85 &  0.029& 0.015 &&\\
\hline
\end{tabular}
\\
\smallskip
\end{center}
%$^a$ Alternative ID: \object{CD-34 510}.	\\
%$^b$ High-velocity Star.\\
%$^c$ Alternative ID: \object{2MASS J01294200-3015464}.\\
%$^a$ Stellar photometry is obtained from the Simbad
%database\footnote{\url{http://simbad.u-strasbg.fr/simbad/}}. Reddening values
%are derived from the \citet[][]{sfd98} reddening maps.\\
%$^e$ Alternative IDs: \object{2MASS J00120998-1756177}, \object{HR37}.
%\\
\smallskip
\end{table*}				  
%\end{sidewaystable}
%#  38.0  &0.7& 2.4     & 1.3
%# -15.8  &0.5& 1.2     & 2.1
%#   3.9  &0.6& $<$1.7  & ---
%#   7.7  &0.6& 3.5     & 1.5
%# 119.8  &0.6& 2.4     & 1.2

%#  -6.0  & 0.9  & 1.9 & 1.3 
% & v$_{helio}$  & $\epsilon(v_{helio})$ & v sin i  & $\epsilon( v sin i)$

%% file: param.tex
\begin{table*}

\caption{Program stars atmospheric parameters. The spectra signal-to-noise
ratios are also indicated, as well as the measured radial velocities.}

\label{PA}
\begin{center}
\begin{tabular}{lcccccrr}
\hline
SPM & \teff         &log g & $\xi$   & A(Fe) & [Fe/H] & S/N  & $v_\mathrm{helio}$  \\
     & K             &      & \kms    &       &        & @670nm &\kms\\
\hline
% ID2    	Teff  logg   vt    Fe	  [Fe/H] 
\object{Sun}   &5777 &4.44 &0.80 &7.50 &       &            &                        \\
\hline
\object{142173} &4330 &1.50 &1.35 &6.81 & -0.69 & 118 	     &  38.0  $\pm$0.7		 \\
\object{171877} &3930 &1.10 &1.40 &6.73 & -0.77 & 109        & -15.8  $\pm$0.5		 \\ 
\object{225245} &3920 &0.65 &1.50 &6.51 & -0.99 & 110        &   3.9  $\pm$0.6		 \\ 
\object{313132} &4530 &2.00 &1.20 &7.51 & +0.01 &  42 	     &   7.7  $\pm$0.6		 \\
\object{343555} &4530 &2.25 &1.00 &6.88 & -0.62 & 106        & 119.8  $\pm$0.6		 \\ 
\hline
\object{HD\,787}$^a$  &4000 &1.50 &1.50 &7.50 & +0.00 & $>$150   &  -6.0  $\pm$ 0.9              \\ 
\object{HD\,787}      &3870 &0.85 &1.27 &7.54 & +0.04 & $>$150   &  -6.0  $\pm$ 0.9              \\ 
\hline
\end{tabular}
\end{center}

$^a$ Atmospheric parameters are assumed from \citet[][]{castilho00}.\\

\end{table*}

%% file: ab.tex
\begin{table*}

\caption{Measured abundances for the program stars. The measured stellar
rotational velocities are also indicated.}\label{Ab}

\begin{center}
\begin{tabular}{lcccccrccccl}
\hline
SPM             & [Fe/H] & [C/Fe] &[N/Fe] &[O/Fe] & C/O & $^{12}$C/$^{13}$C & A(Li)   &   A(Li)    &  A(Li)$_{NLTE}^a$   & A(Li)$_{NLTE}^a$  & $v$\,sin\,$i$ \\
                     &  &      &      &     &		       & &670.8nm &   610.3nm   &  670.8nm &   610.3nm  &\kms\\
\hline
% ID2    	 Fe/H C       N      O      C/O    C12/C13  A(Li)w670 A(Li)w610 
\object{Sun}&  7.50 & 8.49 &  7.95  &  8.83 &       &            &      &      &          &          &\\
\hline
\object{142173} & -0.69 &+0.02 & +0.02  & +0.20 & +0.30 &  $>$15     & 2.80 & 2.52    & 2.73	   &	2.72 & 2.2 $\pm$ 0.7 \\%,~15
\object{171877} & -0.77 &-0.23 & +0.42  & +0.34 & +0.12 &  $>$15     & 2.49 & 2.29    & 2.46	   &	2.52 & $<$1.8		\\   
\object{225245} & -0.99 &-0.44 & +0.54  & +0.08 & +0.14 &      8     & 2.90 & 2.71    & 2.83	 &    2.93 & $<$1.1	       \\  
\object{313132} & +0.01 &+0.07 & +0.01  & -0.07 & +0.63 &     13     & 3.45 & 3.35    & 3.37	 &    3.56 & 3.3 $\pm$ 0.5 \\
\object{343555} & -0.62 &+0.09 & -0.08  & +0.19 & +0.36 &  $>$10     & 1.79 & $<$1.94 & 1.94     &    $<$2.11 & 2.5 $\pm$ 0.8 \\    
\hline
%						   --- [C,N,O/Fe]=-0.24,+0.33,+0.02
\object{HD\,787}$^b$ & +0.00 &  -0.24 &+0.33 &+0.02  & 0.21 &     15  &  2.03 & 2.14 & 2.09      & 2.40 & \\	    
\object{HD\,787}     & +0.04 &  -0.38 &+0.09 &-0.22  & 0.27 &     15  &  1.77 & 2.12 & 1.98      & 2.40 & \\	    
%						   --- [C,N,O/Fe]=-0.38,+0.09,-0.22
\hline
\end{tabular}
\end{center}

$^a$ NLTE corrections are calculated according to \citet[][]{lind09}. Owing to the
grid boundaries, we assume: $\xi$=2.0\kms for all stars; \teff=4000\,K for
\#171877,  \#225245, and HD\,787 (2nd instance); log\,g=1.0 for \#225245 and
HD\,787 (2nd instance); [Fe/H]=0.0 for \#313132 and HD\,787 (2nd instance).\\
$^b$ Atmospheric parameters are assumed from \citet[][]{castilho00}; C, N and O
abundances are taken from \citet[][]{melendez08}. Carbon isotopic ratio is taken
from \citet[][]{dasilva95}. 

\end{table*}

%% file: errors.tex
\begin{table*}

\caption{Estimated errors on absolute abundances and abundance ratios for the
hottest (\#313132) and coolest (\#225245) stars in our sample.}

\label{ER}
\begin{center}
\begin{tabular}{lccccc}
\hline
$\Delta$(El) & S/N & $\Delta$T$_{\rm eff}$=100& $\Delta$log(g)=0.20& $\Delta\xi$=0.10 km/s & $\sigma$$_{\rm tot}$\\
\hline
$\Delta$[Fe/H]$_{\#313132}$ &   0.02&   +0.04&   +0.02&  -0.05 & 0.07\\
$\Delta$[Fe/H]$_{\#225245}$ &   0.01&   +0.01&   +0.03&  -0.05 & 0.06\\
\hline
$\Delta$[C/H]$_{\#313132}$ &    0.03&   -0.02&   +0.04&  +0.00 & 0.05\\
$\Delta$[C/H]$_{\#225245}$ &    0.05&   -0.03&   +0.04&  +0.01 & 0.07\\
\hline 
$\Delta$[N/H]$_{\#313132}$ &    0.05&   -0.01&   +0.12&  -0.01 & 0.13\\
$\Delta$[N/H]$_{\#225245}$ &    0.08&   -0.10&   +0.13&  +0.00 & 0.18\\
\hline
$\Delta$[O/H]$_{\#313132}$ &    0.05&   +0.01&   +0.09&  +0.00 & 0.10\\
$\Delta$[O/H]$_{\#225245}$ &    0.02&   +0.02&   +0.06&  +0.00 & 0.07\\
\hline
$\Delta$($^{12}$C/$^{13}$C)$_{\#313132}$ &  5&0&  0&    0 & 5\\
$\Delta$($^{12}$C/$^{13}$C)$_{\#225245}$ &  2&0&  0&    0 & 2\\
\hline
$\Delta$(log$\epsilon$ Li$_{670}$)$_{\#313132}$ &  0.10& +0.13&  +0.03&  -0.02 & 0.17\\
$\Delta$(log$\epsilon$ Li$_{670}$)$_{\#225245}$ &  0.08& +0.15&  +0.03&  -0.05 & 0.18\\
\hline
$\Delta$(log$\epsilon$ Li$_{610}$)$_{\#313132}$ &  0.03& +0.10&  +0.00&  +0.00 & 0.10\\
$\Delta$(log$\epsilon$ Li$_{610}$)$_{\#225245}$ &  0.02& +0.13&  -0.01&  +0.00 & 0.13\\
\hline
\hline
$\Delta$[C/Fe]$_{\#313132}$ &   0.03&   -0.06&   +0.02&  +0.05 & 0.09\\
$\Delta$[C/Fe]$_{\#225245}$ &   0.05&   -0.04&   +0.01&  +0.06 & 0.09\\
\hline 
$\Delta$[N/Fe]$_{\#313132}$ &   0.05&   -0.05&   +0.10&  +0.04 & 0.13\\
$\Delta$[N/Fe]$_{\#225245}$ &   0.08&   -0.11&   +0.10&  +0.05 & 0.18\\
\hline
$\Delta$[O/Fe]$_{\#313132}$ &   0.05&   -0.03&   +0.07&  +0.05 & 0.10\\
$\Delta$[O/Fe]$_{\#225245}$ &   0.02&   +0.01&   +0.03&  +0.05 & 0.06\\
\hline
\end{tabular}
\end{center}
\end{table*}

%% file: distance_new.tex
\begin{table*}
\caption{Derived distances and orbital parameters of the program stars.}
\label{dist}
\begin{center}
\tiny
\begin{tabular}{lccccccrrrrr}
\hline
 SPM    &   R$_\odot$  &    X	  &   Y   &  Z        &   U	  &	 V   &    W    & e  &  Z$_{max}$ & R$_{p}$ & R$_{a}$	  \\
       &   kpc       & kpc	  & kpc   &  kpc      &  \kms	  &\kms      &   \kms  &    &  kpc    &   kpc & kpc  \\ 
\hline
142173 &  3.6$\pm$0.3	 & 7.92  &   0.47  & -3.48   &-195$\pm$36 &   41$\pm$39 & 36$\pm$9 & 0.83$\pm$0.02 & 3.42$\pm$0.19& 1.13$\pm$0.01&  12.41$\pm$0.14\\ 
171877 &  3.1$\pm$0.3	 & 8.25  &   0.13  & -3.13   &  17$\pm$28 &  -82$\pm$37 & 27$\pm$3 & 0.61$\pm$0.01 & 2.19$\pm$0.06& 2.00$\pm$0.01&   8.27$\pm$0.06\\ 
225245 &  7.3$\pm$0.7	 & 8.55  &   0.10  & -7.28   &  56$\pm$41 &  -52$\pm$43 &  6$\pm$1 & 0.76$\pm$0.03 & 5.11$\pm$0.22& 1.60$\pm$0.02&  11.66$\pm$0.22\\ 
313132 &  0.7$\pm$0.1	 & 8.51  &   0.12  & -0.70   &  50$\pm$11 & -165$\pm$11 & 12$\pm$2 & 0.28$\pm$0.01 & 0.67$\pm$0.01& 5.10$\pm$0.01&   9.09$\pm$0.01\\ 
343555 &  1.1$\pm$0.1	 & 8.60  &   0.14  & -1.07   &  33$\pm$ 9 &  -29$\pm$20 &-84$\pm$3 & 0.88$\pm$0.04 & 1.93$\pm$0.10& 0.46$\pm$0.03&   7.37$\pm$0.15\\ 
\hline
\end{tabular}
\end{center}
\end{table*}